\documentclass[12pt]{iopart}

\usepackage{graphicx}
\usepackage{braket}
\usepackage{bm}
\usepackage{amssymb}

\newcommand\ii{\mathrm{i}}
\newcommand\ifrac[2]{{#1}\mathop{/}{#2}}
\newcommand{\abedt}{$\alpha$-(BEDT-TTF)$_2$I$_3$ }
\newcommand\spinor[2]{
  \left(\begin{array}{c}
   #1 \\ #2
  \end{array}\right)}
\newcommand{\sgn}{\mathop{\rm sgn}}
\newcommand{\tmag}{\tilde{v}_0}
\newcommand{\tang}{\phi_t}
\newcommand{\tmparam}{\gamma}

\newcommand{\cpk}{\bm{k}_0}

\newcommand{\diff}{\mathrm{d}}

\begin{document}

\title{Pressure effects on Dirac fermions in \abedt}

\author{Takahiro Himura, Takao Morinari, and Takami Tohyama}
\address{Yukawa Institute for Theoretical Physics, Kyoto University, Kyoto 606-8502, Japan}
\eads{\mailto{himura@yukawa.kyoto-u.ac.jp}}

\begin{abstract}
We investigate the pressure effect on the layered Dirac fermion system,
which is realized in quasi-two-dimensional organic compound \abedt.
The trajectory of the contact points is investigated using the tight-binding model
with the transfer integrals determined by X-ray diffraction experiments.
Vanishing of the Dirac fermion spectrum, opening of the gap, and pressure dependence of inter-layer magnetoresistance are discussed.
\end{abstract}
\maketitle

\section{Introduction}
\label{intro}

Layered organic conductors, BEDT-TTF [bis(ethyle-nedithiolo)tetrathiofulvalene] salts,
 exhibit various electronic states due to electronic correlation,
 for example, superconductivity, Mott insulator, and charge ordering,
 under variation of pressure or temperature \cite{ishiguro_organic_1998,seo_toward_2004}.

Recently, Katayama {\it et al.} have theoretically suggested that
 the 3/4-filled \abedt salt is a zero-gap system
 under the uniaxial pressure along the BEDT-TTF molecule stack axis ($a$-axis) \cite{kobayashi_superconductivity_2004,katayama_pressure-induced_2006}.
In the zero-gap state, the Fermi surface is reduced to two points,
and the valence and conduction bands contact at two points ($\pm \cpk$),
 which are not located in highly symmetric points in the two-dimensional (2D) Brillouin zone.
Those points shift their positions toward the center of the Brillouin zone
 with increasing pressure \cite{kobayashi_massless_2007,goerbig_tilted_2008,montambaux_merging_2009,montambaux_universal_2009}.
In the vicinity of each contact point,
 two bands show linear dispersion with tilted cone like shape.
In consequence of the cone like dispersion,
 the low-lying excitation properties of the conduction electrons
 are described by a tilted massless Dirac equation \cite{kobayashi_massless_2007}.
The first principles calculations support this Dirac cone structure \cite{ishibashi_ab_2006,kino_first-principles_2006}.

The massless Dirac electron system has been also found in graphene which is a single layer of graphite \cite{novoselov_electric_2004,novoselov_two-dimensional_2005,zhang_experimental_2005}.
However, in contrast to graphene, \abedt is not a two dimensional system but a multi-layered bulk material.
In \abedt crystal,
 conducting layers of BEDT-TTF molecules
 and insulating layers of I$_3$ anions
 stack alternatively along the c-axis.
Since the conductive layers are separated by the insulating layers,
 interlayer transfer energy is sufficiently small
  and thus this system has a strong 2D nature.

It has theoretically been explained by Osada \cite{osada_negative_2008}
 that the experimentally observed negative interlayer magnetoresistance \cite{tajima_effect_2009} is due to the zero-mode Landau Level of the massless Dirac fermion.
 In addition, we have proposed that the presence of a tilted and anisotropic Dirac cone can be verified using the interlayer magnetoresistance \cite{morinari_possible_2009}.
The interlayer magnetoresistance in \abedt depends on the in-plane magnetic field direction because of tilt.

The hopping parameters of the \abedt system are controlled by pressure.
Two Dirac cone locations in the Brillouin-zone also move by pressure,
therefore the distance between valleys would be changed.
This feature is very intriguing, because the inter-valley scattering effect would be affected by the change of Dirac cone locations.

In this paper, we investigate the pressure effect on the trajectory of the two Dirac points by changing transfer energies within tight-binding model.
We also calculate the pressure effect on the interlayer magnetoresistance.
We use the parameters of the Weyl equation estimated by the tight-binding model \cite{kobayashi_massless_2007,katayama_pressure-induced_2006}
 with the transfer integrals determined by X-ray diffraction experiments \cite{kondo_crystal_2005}, and discuss the Dirac fermion merging and the gap opening under high pressure region.

The organization of this paper is as follows.
In section~2, we calculate the pressure dependence of the Dirac cone parameters in \abedt by the tight-binding model.
In section~3, we discuss the merging behavior of Dirac points.
In section~4, we show the exact solution of the Landau level on the tilted Weyl equation.
In section~5, we calculate the pressure dependence of the interlayer magnetoresistance by using the parameters estimated from the tight-binding model.
Section~6 gives conclusions of this work.

\section{Pressure dependence}
\label{pressure-dependence}

\begin{figure}
  \begin{center}
    \includegraphics[width=0.8\linewidth]{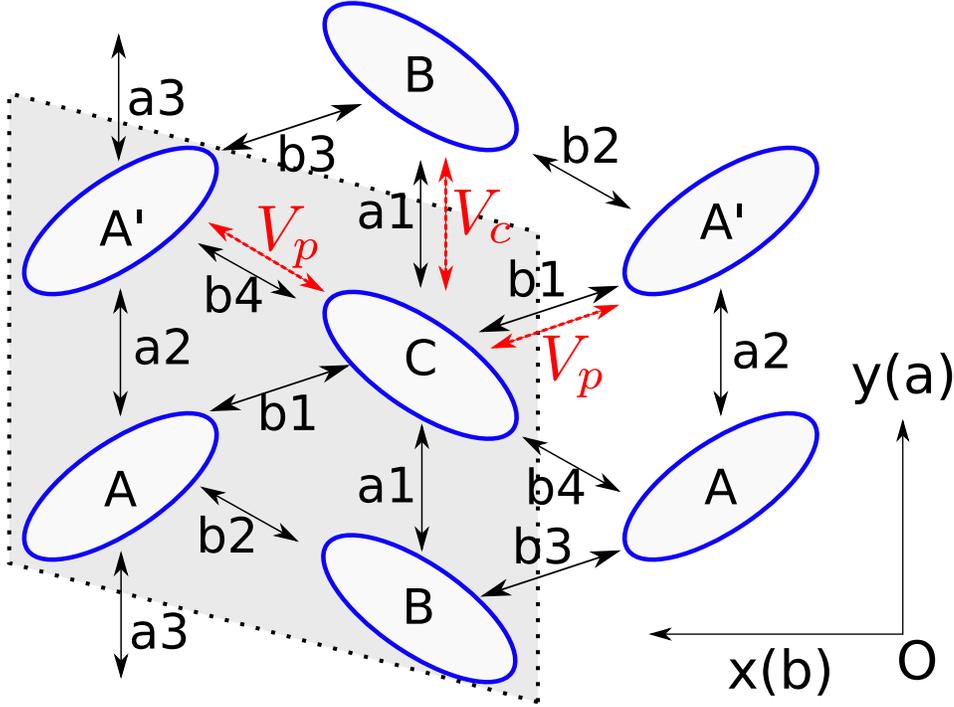}
    \caption{The conducting layer structure of \abedt, where the unit cell is given by the shadowed region.
      The bonds labeled by a1,$\cdots$,a3,b1,$\cdots$,b4 have transfer energy. $V_c$ ($V_p$) represents repulsive Coulomb interaction along a(b)-axis}
    \label{fig:molecular-structure}
  \end{center}
\end{figure}

In \fref{fig:molecular-structure}, the basic model describing electronic state in $\alpha$-(BEDT-TTF)$_2$I$_3$ is shown \cite{kobayashi_superconductivity_2004,kobayashi_superconductivity_2005,katayama_pressure-induced_2006,kobayashi_massless_2007,kobayashi_theoretical_2009}.
The unit cell consists of four BEDT-TTF molecules named by A, A', B and C according to charge disproportionation.
To consider the Coulomb interaction between molecules,
we use the extended Hubbard model which is given by
\begin{equation}\eqalign{
  H = & \sum_{(i \alpha; j \beta), \sigma} t_{i \alpha; j \beta} a_{i \alpha \sigma}^\dagger a_{j \beta \sigma}
  + \sum_{i \alpha} U a_{i \alpha \uparrow}^\dagger a_{i \alpha \downarrow}^\dagger a_{i \alpha \downarrow} a_{i \alpha \uparrow} \\
  & + \sum_{(i \alpha; j \beta), \sigma, \sigma'} V_{\alpha\beta} a_{i \alpha \sigma}^\dagger a_{j \beta \sigma'}^\dagger a_{j \beta \sigma'} a_{i \alpha \sigma},
  \label{abedt-tight-binding-hami}}
\end{equation}
where $i$ and $j$ denote indices of the unit cell,
and $\alpha$ and $\beta$ are indices of BEDT-TTF molecules in the unit cell.
In the first term, $a_{i \alpha \sigma}^\dagger$ ($a_{i \alpha \sigma}$) denotes the creation (annihilation) operator for the electron of spin $\sigma \ (= \uparrow, \downarrow)$ at the $i$th site,
and $t_{i \alpha; j \beta}$ is the transfer energy between the $(i, \alpha)$ and $(j, \beta)$ sites.
The second and the last terms denote repulsive Coulomb interactions
where $U$ is the on-site interaction,
and $V_{\alpha\beta}$ the anisotropic nearest-neighbor interaction.
Following Refs.\cite{kobayashi_massless_2007,katayama_pressure-induced_2006},
we introduce the effect of the uniaxial pressure along the $a$-axis ($P_a$)
by changing the transfer energies $t_A$
\begin{equation}
  t_A(P_a) = t_A (0) (1 + K_A P_a).
\end{equation}
The transfer energies $t_A$ and coefficients $K_A$ $(A = a1, a2, \cdots, b4)$ are obtained from
the data at $P_a = 0$ kbar and at $P_b = 2$ kbar \cite{kondo_crystal_2005}.
We note that the linear variation of the
transfer energies under uniaxial pressure is reasonable for weak
pressure.  At high pressure, there are deviations from the linear dependence.
Since details of pressure dependence of the transfer energies,
especially at high pressure, are not known,
we use the linear functional forms for simplicity.
Although critical pressures below cannot be taken seriously,
the purpose of the present study is to demonstrate pressures effects on
magnetoresistance and merging effects of Dirac fermions.
The Coulomb interactions $U, V_{\alpha \beta}$ are treated within the Hartree approximation.
The mean field Hamiltonian is given by
\begin{equation}\eqalign{
  H_{\textrm{MF}} & = \sum_{\bm{k} \alpha \beta \sigma} \epsilon_{\alpha \beta \sigma}(\bm{k}) a_{\bm{k} \alpha \sigma}^\dagger a_{\bm{k} \beta \sigma} \\
  & \quad - \sum_{\alpha} U_\alpha \braket{n_{\alpha\uparrow}} \braket{n_{\alpha\downarrow}}
  - \sum_{{<\alpha, \beta> \atop \sigma, \sigma'}} V_{\alpha\beta} \braket{n_{\alpha\sigma}} \braket{n_{\beta\sigma'}},\\
  \epsilon_{\alpha \beta \sigma}(\bm{k})
  & = \delta_{\alpha\beta}
  [
  U_\alpha \braket{n_{\alpha\bar{\sigma}}} + \sum_{\beta' \sigma'} V_{\alpha\beta'}\braket{n_{\beta'\sigma'}}
  ] + \sum_{\bm{\delta}} t_{\alpha\beta} \e^{\ii \bm{k}\cdot\bm{\delta}},
  \label{abedt-mean-field-hami}}
\end{equation}
where $a_{\bm{k} \alpha \sigma}$ is the Fourier transform of $ a_{i \alpha \sigma}$,
$\braket{n_{\alpha \sigma}}$ is the averaged number of electrons $(1/N) \sum_{i} \braket{a_{i\alpha\sigma}^\dagger a_{i\alpha\sigma}}$,
$\bar{\sigma}$ denotes the inversion of the spin $\sigma$,
and $\bm{\delta}$ denotes the vector connecting the nearest neighbor sites of the unit cell.

The Hamiltonian \eref{abedt-mean-field-hami} is diagonalized as
\begin{equation}
  \sum_{\beta=1}^4 \epsilon_{\alpha\beta\sigma}(\bm{k}) d_{\beta r \sigma}(\bm{k})
  = \xi_{r \sigma} (\bm{k}) d_{\alpha r \sigma}(\bm{k}),
\end{equation}
where $r$ is an index of eigenstates with eigenvalues $\xi_{r \sigma}$,
which are arranged by a descending order $\xi_{1 \sigma}(\bm{k}) \ge \xi_{2 \sigma}(\bm{k}) \ge \xi_{3 \sigma}(\bm{k}) \ge \xi_{4 \sigma}(\bm{k})$.
$d_{\alpha r \sigma}(\bm{k})$ is the corresponding eigenvector.

The averaged number of electrons $\braket{n_{\alpha \sigma}}$
is written by
\begin{equation}
  \label{self-consistent-eq}
  \braket{n_{\alpha\sigma}} = \sum_{\bm{k}} \sum_{r = 1}^4
  \frac{|d_{\alpha r \sigma}(\bm{k})|^2}{\exp[(\xi_{r\sigma}(\bm{k}) - \mu)/ k_B T] + 1},
\end{equation}
where $T$ is a temperature and $k_B$ denotes the Boltzmann constant.
The chemical potential $\mu$ is determined by the condition $\frac{3}{4} = \frac{1}{8} \sum_{\alpha \sigma} \braket{n_{\alpha\sigma}}$,
because of 3/4-filling.
The parameters $U = 0.4$ eV, $V_c=0.17$ eV, and $V_p = 0.05$ eV are chosen \cite{kobayashi_superconductivity_2005,kobayashi_massless_2007}.

In this paper, we consider the zero-gap state.
The conduction band ($r = 1$) and the valence band ($r = 2$) are degenerate at the two points $\bm{k}_0$ and $- \bm{k}_0$,
and in the vicinity of the contact point $\bm{k}_0$,
the Hamiltonian is written by \cite{kobayashi_massless_2007}
\begin{equation}
H_\tau(\tilde{\bm{k}}) = \tau \left( v_0^x \tilde{k}_x + v_0^y \tilde{k}_y \right) \sigma_0 + \tau v_x \tilde{k}_x \sigma_x + v_y \tilde{k}_y \sigma_y,
\label{tilted-hami}
\end{equation}
where
$\tilde{\bm{k}} = \bm{k} - \bm{k}_0$,
$\sigma_x$ and $\sigma_y$ are Pauli matrices and $\sigma_0$ is the identity matrix, $\tau = \pm$ denotes the valley index which corresponds with $\bm{k}_0$ and $- \bm{k}_0$ respectively.
These contact points occur in pairs and can be described by
independent degrees of freedom, which leads to a twofold valley degeneracy.
The valley degree of freedom is called as ``valley spin''.
We note that the $k_x$- and $k_y$-axes are not along the crystalline a$^\ast$- and b$^\ast$-axes, respectively
(the superscript~$^\ast$ means the reciprocal), because the system is rotated in order to remove complexity of Hamiltonian.
We define the angle made by $k_x$ and b$^\ast$ as $\phi_0$.

For convenience, we define the Dirac cone parameters as
\begin{equation}
\tmag \e^{\ii \tang} \equiv \frac{v_0^x}{v_x} + \ii \frac{v_0^y}{v_y},
\quad
\alpha \equiv \sqrt{\frac{v_x}{v_y}},
\quad
\gamma \equiv \sqrt{1 - \tilde{v}_0^2},
\label{define_v_varphi}
\end{equation}
where $\tmag$ and $\tang$ represent the tilting magnitude and the direction of the Dirac cone, respectively.
The parameter $\alpha$ represents the strength of anisotropy coming from non-tilting effect.
The parameter $\gamma$ measures the strength of tilt of the Dirac cone which satisfies the relation $0 < \gamma \le 1$ ($\gamma = 1$ for non-tilting case).

\begin{figure}[htb]
  \begin{center}
    \includegraphics[width=0.9\linewidth]{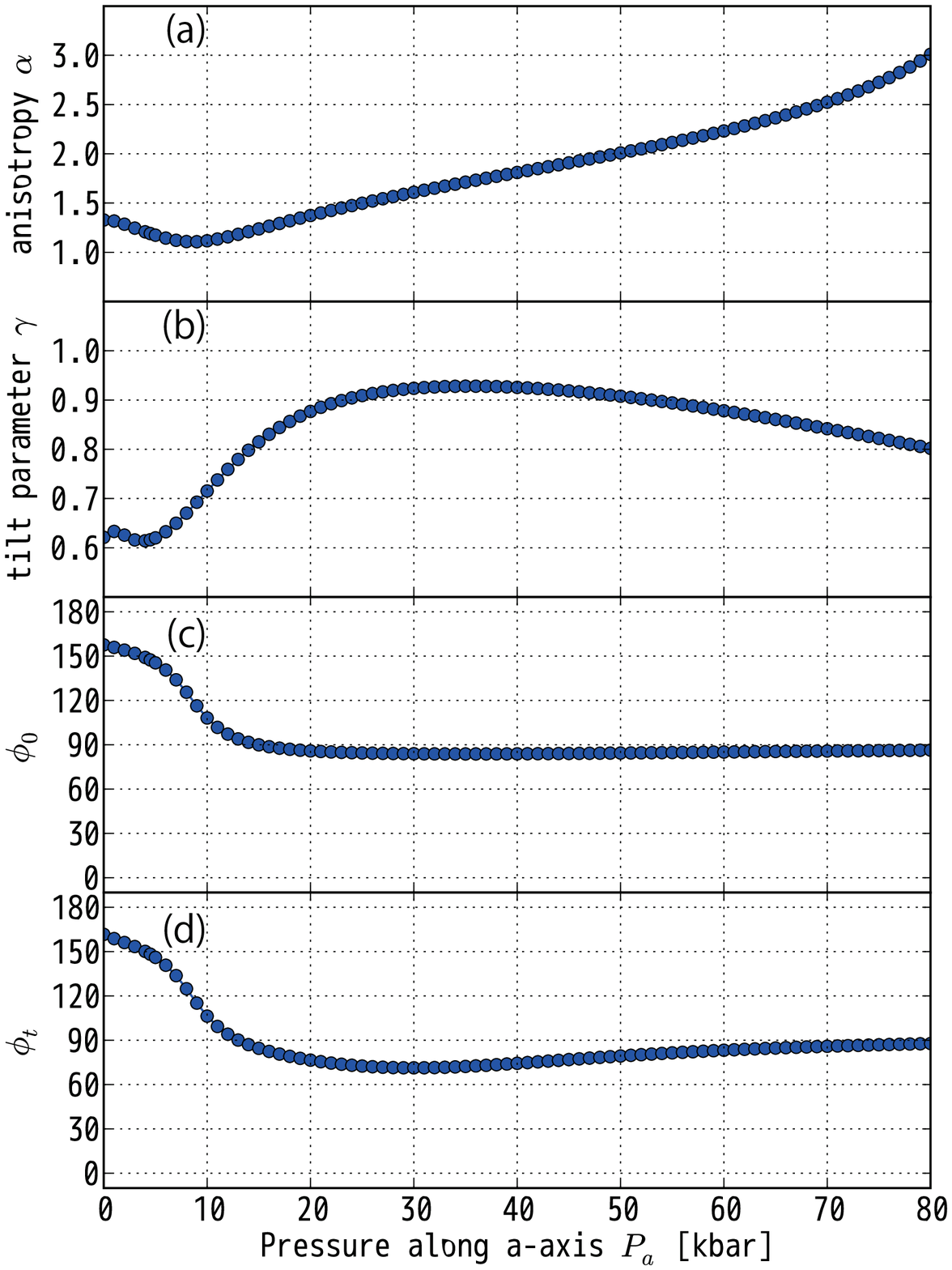}
   \caption{Pressure dependence of the Dirac cone parameters for $U = 0.4$ eV, $V_c=0.17$ eV, and $V_p = 0.05$ eV.
     (a) The anisotropy $\alpha$, (b) the amplitude of tilt $\gamma$ ($\gamma = 1$ for non-tilting case), (c) the angle $\phi_0$ made by $k_x$ and crystalline b$^\ast$ axis, and (d) the direction of tilt $\phi_t$.}
    \label{fig:parameter}
  \end{center}
\end{figure}

\Fref{fig:parameter} shows the pressure dependence of the Dirac cone parameters under the uniaxial pressure $P_a$.
\Fref{fig:parameter}(a) shows the pressure dependence of the anisotropy coming from non-tilting effect.
At $P_a \approx 8$ kbar, the system is almost isotropic ($\alpha \approx 1$) 
because the hopping parameter $t_{a2}$ takes almost the same value as $t_{b1}$ and $t_{b2}$.
At $P_a > 10$ kbar, $\alpha$ increases with pressure.
This growth results in the increase of the interlayer magnetoresistance peak with respect to azimuthal angle dependence.
In the region $4.5 \leq P_a < 8$ kbar,
$\alpha$ decreases with increasing pressure.
\Fref{fig:parameter}(b) shows the pressure dependence of the amplitude of Dirac-cone tilting.
At $P_a \approx 5$ kbar, the tilt of Dirac cone takes maximum.
In the region $5 < P_a < 35$ kbar, $\gamma$ increases with increasing pressure,
thus, the tilt of Dirac cone decreases.
At $P_a > 35$ kbar, the tilt of Dirac cone increases again with pressure.
\Fref{fig:parameter}(c) shows the angle made by $k_x$ and crystalline b$^\ast$-axis.
In the high pressure region, $P_a \ge 20$ kbar,
this angle becomes almost constant and the $k_x$ axis is parallel to the crystalline b$^\ast$-axis.
We recall that the pressure $P_a$ is uniaxial.
In the high-pressure region,
the transfer integrals $t_{a1}$ and $t_{a2}$ are enhanced by the uniaxial pressure,
so the energy contour becomes elliptic and shrinks along the a$^\ast$-axis.
\Fref{fig:parameter}(d) shows the azimuthal angle of the tilting direction.
In the high pressure region, $P_a > 30$ kbar,
the tilt of Dirac cone is almost along the $k_y$-axis.

\section{Merging Dirac points}

\begin{figure}[htb]
  \begin{center}
    \includegraphics[width=0.95\linewidth]{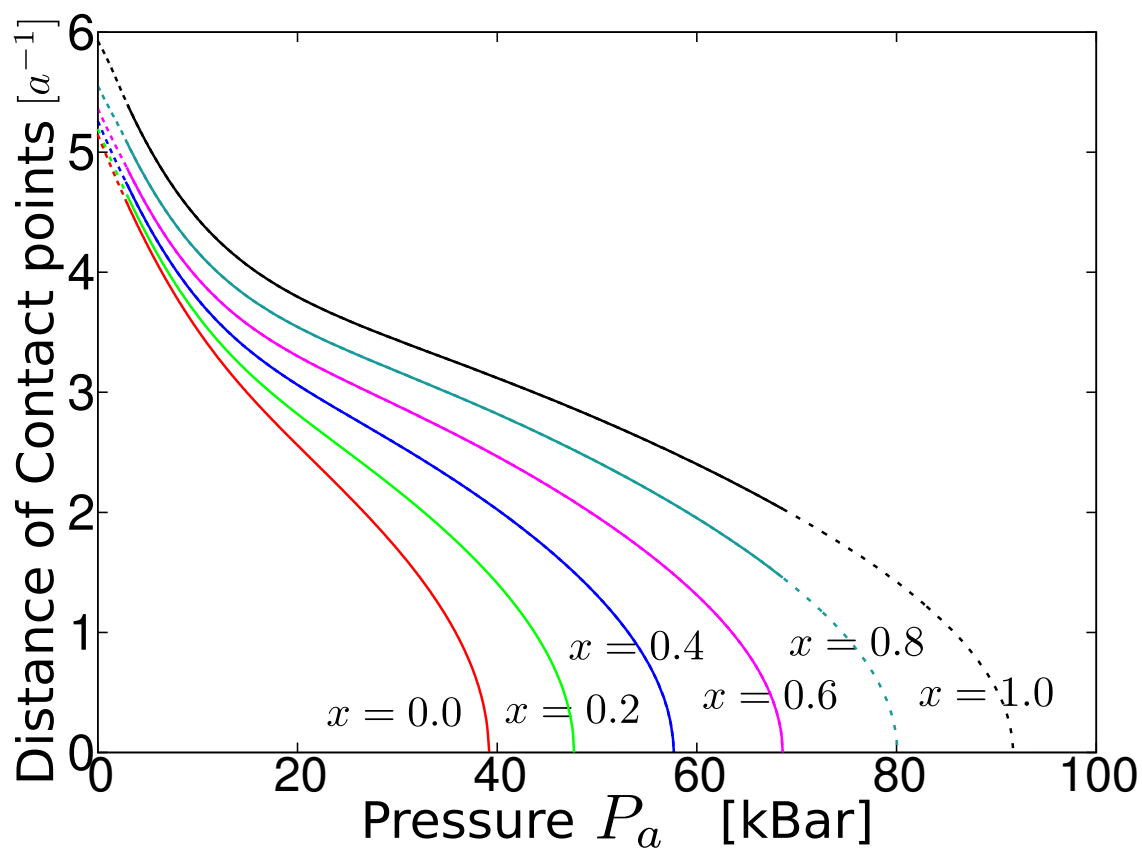}
    \caption{Pressure dependence of the distance between two contact points. The interaction is taken to be $U = 0.4x$, $V_c = 0.17x$, and $V_p = 0.05x$.
      The dashed line denotes that the contact points is not located on the Fermi level.}
    \label{fig:contact_point_distance}
  \end{center}
\end{figure}
\begin{figure}[htb]
  \begin{center}
    \includegraphics[width=0.9\linewidth]{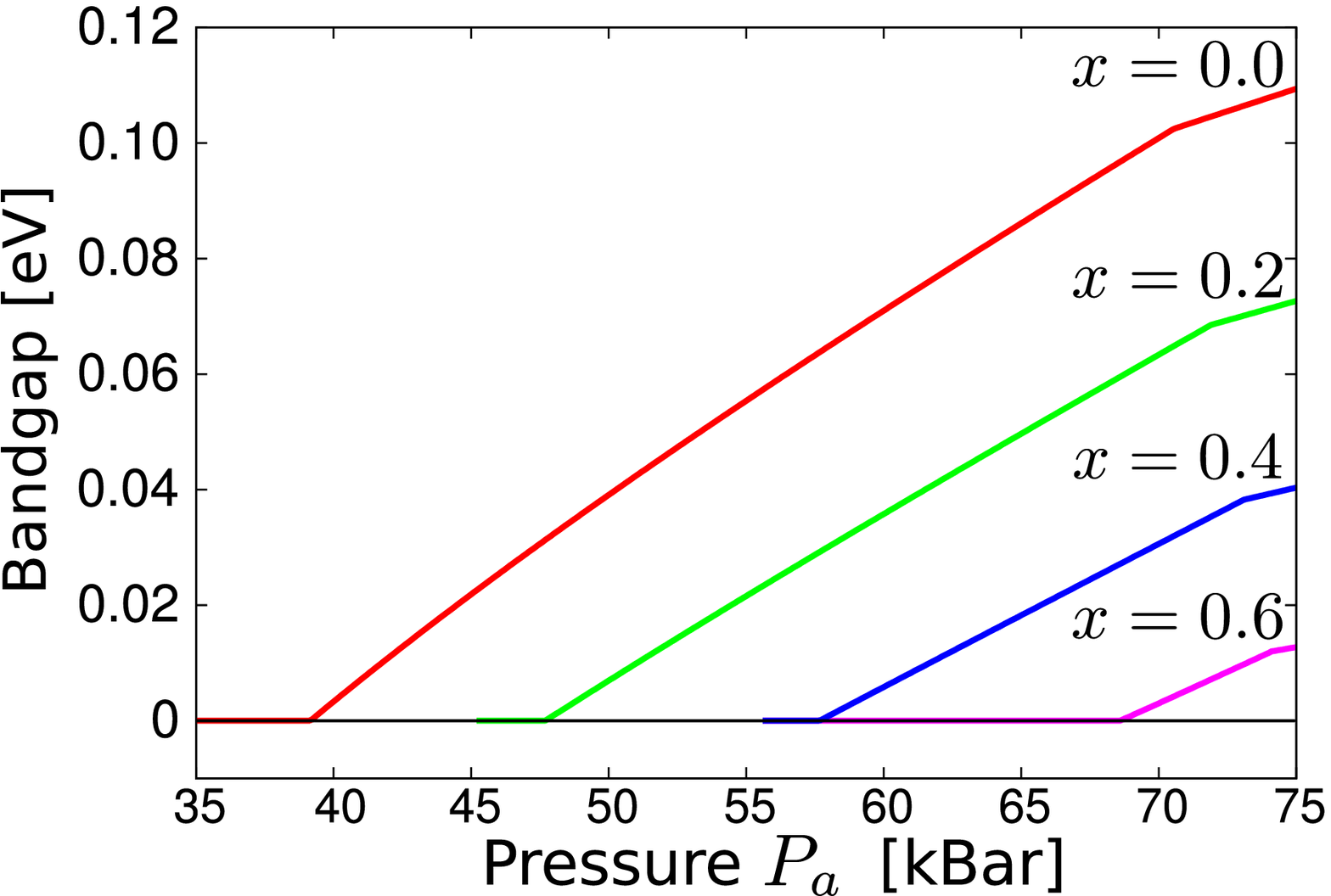}
    \caption{Pressure dependence of bandgap. The interaction parameters are taken in the same manner as \fref{fig:contact_point_distance}}
    \label{fig:bandgap}
  \end{center}
\end{figure}

As pressure increases, the two contact points approach each other, and then they merge into the single point.
After merging, the contact points vanish and the gap opens between the electron and hole bands.
Montambaux {\it et al.} have proposed the universal 2$\times$2 Hamiltonian to describe the motion and merging behavior of Dirac points, and they have obtained a semiclassical description of the Landau levels spectrum \cite{goerbig_tilted_2008,montambaux_merging_2009,montambaux_universal_2009}.
This model describes continuously the Landau level coupling between valleys associated with two Dirac points in the vicinity of the merging Dirac points.

Here we calculate the trajectory and the gap opening behavior of the Dirac points by the 4-band tight-binding model described in section~\ref{pressure-dependence}.
\Fref{fig:contact_point_distance} shows the pressure dependence of the distance of the Dirac points for several values of $U$, $V_c$, and $V_p$ scaled by an interaction parameter $x$.
The contact points exist on wide pressure range.
However, in some pressure region, they are not located at the Fermi level.
Under low pressure regions, $P_a < 3$ kbar, the contact points exist but they are not located at the Fermi level because of the existence of the hole and electron pockets, which are denoted by dashed lines in \fref{fig:contact_point_distance}.
Under high pressure region, $P_a > 68$ kbar,
the contact points in the case of the interaction parameter $x = 0.8$ and $x = 1.0$,
are not located at the Fermi level as denoted by dashed line in \fref{fig:contact_point_distance}.
In the case of the interaction parameter $0 \le x \le 0.6$,
the contact points are located just on the Fermi level
in vicinity of the critical pressure for the merging of the Dirac points.
After merging, the contact points vanish and the gap opens between the electron and hole bands.
\Fref{fig:bandgap} shows the pressure dependence of the gap between the two subbands.
The bandgap depends on the pressure linearly.
The critical pressure for the merging of the Dirac points increases with increasing the parameter $x$.
This critical pressure increase is not general behavior for other interaction parameters.
In this case, the interactions $U$, $V_c$ and $V_p$ are taken so that the charge disproportionation pattern becomes
the stripe pattern which is consistent with the experiment.
We do not understand the
mechanism of this upward shift. But this upward shift suggests that
the Dirac fermions are stabilized by increasing the interaction
parameters. Probably this is associated with the enhancement of charge
disproportionation. We would like to investigate this point in a
future publication.

\section{Exact solution of Landau level on tilted Weyl equation}

As shown in the section~\ref{pressure-dependence}, Dirac fermions in the \abedt system are described by tilted Weyl equation.
Reflecting the tilt of the Dirac cone, the Landau level wave functions are anisotropic.
In this section we derive the exact Landau level wave functions of those Dirac fermions under magnetic field.

First, we rescale the system to remove the anisotropy coming from non-tilting effects:
$v_x \pi_x \rightarrow v \pi_x, \  v_y \pi_y \rightarrow v \pi_y$
where $v = \sqrt{v_x v_y}$ and $\pi_i = - \ii \hbar \partial_{x_i} + e A_i \, (i = x,y)$.
Second,
we rotate the system by the angle $\tang$ in the plane so that the tilting direction of the Dirac cone would be along with rotated $k_x$ axis:
\begin{equation}
\spinor{\pi_x}{\pi_y}
=
\left(\begin{array}{cc}
  \cos{\tang} & - \sin{\tang}
  \\
  \sin{\tang} & \cos{\tang}
\end{array}\right)
\spinor{\pi_x'}{\pi_y'}.
\label{rotate-system}
\end{equation}
After these transformations, the tilted Weyl Hamiltonian is written as
\begin{equation}
H_\tau(\bm{p}) = v U_\tau^\dagger(\tang) \left[ \tau \tilde{v}_0 \pi_x' \sigma_0
+ \tau \pi_x' \sigma_x + \pi_y' \sigma_y \right] U_\tau(\tang),
\end{equation}
\begin{equation}
U_\tau(\tang) = \cos(\tang / 2) \sigma_0 + \ii \tau \sin(\tang / 2) \sigma_z.
\end{equation}
We multiply both sides of the Schr\"{o}edinger equation $H \psi = E \psi$ by the operator
$U_\tau^\dagger(\tang) \left[ \tau \pi_x' \sigma_x + \pi_y' \sigma_y \right] U_\tau(\tang)$ from the left,
and then after some algebra we obtain
\begin{equation}\eqalign{
\fl v^2 \left[
 (1 - \tilde{v}_0^2) \left(
   \pi_x' + \frac{\tau}{v} \frac{E \tilde{v}_0}{1 - \tilde{v}_0^2}
 \right)^2 + \pi_y'^2
\right]
\psi
\\
= \left[\frac{E^2}{1 - \tilde{v}_0^2}
- \tau v^2 \frac{\hbar^2}{l_z^2} U_\tau^\dagger(\tang)
\left(\sigma_z + i \tilde{v}_0 \sigma_y \right) U_\tau(\tang) \right] \psi.
\label{schoeredinger-eq}}
\end{equation}
We redefine the momentum operator as
\begin{equation}
\tilde{\pi}_x = \sqrt{\gamma} \left(
 \pi_x' + \frac{\tau}{v} \frac{E \tilde{v}_0}{1 - \tilde{v}_0^2}
\right)
,\quad
\tilde{\pi}_y = \frac{1}{\sqrt{\gamma}} \pi_y'^2.
\label{shift-dynamic-momentum}
\end{equation}
Both $\tilde{\pi}_x$ and $\tilde{\pi}_y$ satisfy the commutation relation $ [\tilde{\pi}_x, \tilde{\pi}_y] = - \ii \ifrac{\hbar^2}{l_z^2}$,
where $l_z = \sqrt{\ifrac{\hbar}{e B_z}}$ is the magnetic length.
We rewrite equation \eref{schoeredinger-eq} by the redefined momentum operator
\begin{equation}
  \left[
    \tilde{\pi}_x^2 + \tilde{\pi}_y^2
  \right]
  \psi
  =
  \frac{1}{\gamma} \frac{\hbar^2}{l_z^2}
  \left(\begin{array}{cc}
      \frac{\varepsilon^2}{1 - \tilde{v}_0^2} - \tau & - \tau \tmag \e^{\ii \tau \tang} \\
      \tau \tmag \e^{- \ii \tau \tang} & \frac{\varepsilon^2}{1 - \tilde{v}_0^2} + \tau
    \end{array}\right) \psi,
  \label{shifted-energy-condition}
\end{equation}
where $E = \frac{\hbar}{l_z} v \varepsilon$.
We define the ladder-operator
\begin{equation}
\tilde{a} = \frac{l_z}{\sqrt{2}\hbar} \left( \tilde{\pi}_x - \ii \tilde{\pi}_y \right)
,
\end{equation}
which satisfies the commutation relation $[\tilde{a},\tilde{a}^\dagger] = 1$.
In addition, we define the number operator
$\tilde{N} = \tilde{a}^\dagger \tilde{a} = \frac{l_z^2}{2 \hbar^2} (\tilde{\pi}_x^2 + \tilde{\pi}_y^2) - \frac{1}{2}$.
We take the eigenstate of the number operator,
$\tilde{N} \phi_n = n \phi_n$.
The eigenstate of Hamiltonian \eref{tilted-hami} is denoted by $\phi_n$.

Now we comment on the difference between tilted and non-tilted Dirac cones.
If the Dirac cone is not tilting, the right hand side of equation~\eref{shifted-energy-condition} becomes diagonal.
In the tilted case, the off-diagonal part does not vanish,
hence the wave functions are linear combinations of $\phi_n$ and $\phi_{n+1}$.
We write the wave function as $\psi = (u, v)^T \phi_n$,
and substitute this into equation~\eref{shifted-energy-condition},
then we get the relation
\begin{equation}
\left(\begin{array}{cc}
  \varepsilon^2 - \gamma^2 \tau - \gamma^3 (2 n + 1) & - \tau \gamma^2 \tmag \e^{- \ii \tau \tang} \\
  \tau \gamma^2 \tmag \e^{\ii \tau \tang} & \varepsilon^2 + \gamma^2 \tau - \gamma^3 (2 n + 1) \\
\end{array}\right) \spinor{u}{v} = 0.
\label{wfcondition}
\end{equation}
For this equation to have a solution, the determinant of the left-hand side matrix must be equal zero,
so the eigenenergy $\varepsilon$ becomes
\begin{equation}
\varepsilon = \pm \sqrt{\gamma^3 (2 n + 1 \pm 1)}.
\end{equation}
When $\varepsilon = \pm \sqrt{2 \gamma^3 n}$, the wave functions are given by
\begin{equation}
\psi_n^\tau
=
\frac{1}{\sqrt{2 (1 + \tau \gamma)}} \spinor{- \tau \tmag \e^{- \ii \tau \tang}}{\tau + \gamma} \phi_n.
\label{wave-function-n}
\end{equation}
When $\varepsilon = \pm \sqrt{2 \gamma^3 (n+1)}$, the wave functions are given by
\begin{equation}
\psi_n^\tau
=
\frac{1}{\sqrt{2 (1 + \tau \gamma)}} \spinor{\tau + \gamma}{- \tau \tmag \e^{\ii \tau \tang}} \phi_n.
\label{wave-function-n2}
\end{equation}
Then, the wave function $\psi_n$ which has the eigenenergy $\varepsilon = \sgn(n) \sqrt{2 \gamma^3 |n|}$ reads
\begin{equation}\eqalign{
\psi_n^\tau
& =
\frac{A_n}{\sqrt{2 (1 + \tau \gamma)}} \spinor{- \tau \tmag \e^{- \ii \tau \tang}}{\tau + \gamma} \phi_{|n|} \\
& \quad +
\frac{B_n}{\sqrt{2 (1 + \tau \gamma)}} \spinor{\tau + \gamma}{- \tau \tmag \e^{\ii \tau \tang}} \phi_{|n| - 1}.}
\end{equation}
The coefficients $A_n$ and $B_n$ satisfy the relation $A_n \e^{-\ii \phi} = \sgn(n) B_n$ which is determined from the Sch\"{o}redinger equation $H \psi_n = E_n \psi_n$.

Finally, the energy and eigenstate are written as
\begin{equation}
  E_n = \sgn(n) \sqrt{2 \hbar v_x v_y |e| B_z \gamma^3 |n|},
  \label{landau-level-energy}
\end{equation}
\begin{equation}\eqalign{
  \psi_n^\tau
  & =
  \frac{1}{2 \sqrt{1 + \tau \gamma}} \Biggl[
  \spinor{- \tau \tmag}{(\tau + \gamma) \e^{\ii \tau \tang}} \phi_{|n|}
  \\
  & \quad +
  \spinor{\tau + \gamma}{- \tau \tmag \e^{\ii \tau \tang}} \phi_{|n| - 1}
  \Biggl]
  \quad (n \neq 0),}
\end{equation}
and
\begin{equation}\eqalign{
  \psi_0^\tau
  & =
  \frac{1}{\sqrt{2 (1 + \tau \gamma)}} \spinor{- \tau \tmag \e^{- \ii \tau \tang}}{\tau + \gamma} \phi_0.
\label{full-wave-function}}
\end{equation}

The explicit form of the Landau level wave functions
depends on the choice of the gauge.
In order to see anisotropy of the wave function,
it is convenient to take symmetric gauge.
On the other hand, for the calculation of the inter-layer magnetoresistance,
it is convenient to take the Landau gauge.
Below we show both cases separately.

\subsection{Symmetric gauge case}
\label{symmetric-gauge-case}

As we shall see later, the interlayer magnetoresistance in \abedt depends on the in-plane magnetic field direction because of anisotropy in the Landau level wave function.
In order to get a clear picture, we solve the tilted Weyl equation with the symmetric gauge $\bm{A} = \frac{1}{2} \bm{B} \times \bm{r}$.

The presence of the in-plane magnetic field is taken into account by a gauge-transformation
\begin{equation}
\psi
= \exp\left[
  - \ii \left(
    \frac{x}{l_y^2}
    -
    \frac{y}{l_x^2}
  \right)
  \frac{z}{2}
\right]
\psi',
\end{equation}
with the magnetic length $l_\mu = \sqrt{\ifrac{\hbar}{e B_\mu}} \  (\mu = x,y)$.
After this transformation, the vector potential is given by
$\pi_x = - \ii \hbar \partial_x + \frac{1}{2} e B_z y$
and
$\pi_y = - \ii \hbar \partial_y - \frac{1}{2} e B_z x$.
We rescale and rotate the system as
\begin{equation}\eqalign{
{} & v_x^{-1} x \rightarrow v x, \  v_y^{-1} y \rightarrow v y
,\\
{} & \spinor{x}{y}
=
\left(\begin{array}{cc}
  \cos{\tang} & - \sin{\tang}
  \\
  \sin{\tang} & \cos{\tang}
\end{array}\right)
\spinor{x'}{y'},
\label{rotate-coordinate}}
\end{equation}
respectively.
The transformation of equation~\eref{shift-dynamic-momentum} is equivalent to
$\tilde{x} = \frac{1}{\sqrt{\gamma}} x'$
and
$\tilde{y} = \sqrt{\gamma} \left(y' + \frac{\tau}{v} \frac{E \tilde{v}_0}{\gamma^2}\frac{2}{e B_z}\right)$.
The center coordinate of the cyclotron motion $\tilde{X},\ \tilde{Y}$ reads
\begin{equation}
\tilde{X} = \tilde{x} - \frac{l_z^2}{\hbar} \tilde{\pi}_y, \quad
\tilde{Y} = \tilde{y} + \frac{l_z^2}{\hbar} \tilde{\pi}_x.
\end{equation}
This satisfies the non-vanishing commutation relation as $[\tilde{X}, \tilde{Y}] = \ii l_z^2$,
and wave functions cannot be simultaneously eigenfunctions of both of them.
We choose to use the operator $\tilde{X}^2 + \tilde{Y}^2$ which also commutes with Hamiltonian.
In the symmetric gauge case, this operator is given by
\begin{equation}
\tilde{X}^2 + \tilde{Y}^2 =
2 l_z^2 \left(
  \tilde{N} - \tilde{L}_z + \frac{1}{2}
\right)
\end{equation}
with the number operator $\tilde{N} = \tilde{a}^\dagger\tilde{a}$
, and the angular momentum
$\tilde{L}_z = \frac{1}{\hbar}(\tilde{x}\tilde{p}_y - \tilde{y}\tilde{p}_x) = \hbar (\tilde{a}^\dagger \tilde{a} - \tilde{b}^\dagger \tilde{b})$.
Here, we define the ladder-operator $\tilde{b}$ as
\begin{equation}
\tilde{b} = \frac{1}{\sqrt{2} l_z} \left(\tilde{X} + \ii \tilde{Y}\right).
\end{equation}
$\tilde{a}$ and $\tilde{b}$ satisfy commutation relations:
$[\tilde{b}, \tilde{b}^\dagger] = 1$,
$[\tilde{a}, \tilde{b}] = [\tilde{a}, \tilde{b}^\dagger] = 0$.
We can rewrite these expressions by introducing complex coordinates
$\tilde{\chi} = \frac{\tilde{x} - \ii \tilde{y}}{l_z}$
\begin{equation}
\tilde{a} = \frac{1}{\sqrt{2}} (- \ii \partial_{\tilde{\chi}^\ast} - \ii \frac{1}{2} \tilde{\chi})
, \quad
\tilde{b} = \frac{1}{\sqrt{2}} (\partial_{\tilde{\chi}} + \frac{1}{2} \tilde{\chi}^\ast).
\end{equation}

Using these operators, we find that the eigenstates are denoted by a ket vector $\ket{n,m}$ ($n \geq 0,\, m \leq - n$),
where $\tilde{a}^\dagger \tilde{a} \ket{n,m} = n \ket{n,m}$
, and
$\tilde{b}^\dagger \tilde{b} \ket{n,m} = (n - m) \ket{n,m}$.
The eigenvalue of $\tilde{L}_z$ is $\hbar m$.

The wave function for zero-mode eigenfunction $\ket{0,0}$
is obtained by solving $\tilde{a} \ket{0,0} = \tilde{b} \ket{0,0} = 0$.
In the coordinate representation $\phi_{0,0} (\bm{r}) \equiv \braket{\bm{r}|0,0}$,
\begin{equation}
\phi_{0,0}(\tilde{\bm{r}}) = \frac{1}{\sqrt{2 \pi} l_z} \e^{-\frac{|\tilde{\chi}|^2}{4}}
= \frac{1}{\sqrt{2 \pi} l_z} \exp\left(-\frac{\tilde{x}^2 + \tilde{y}^2}{4 l_z^2}\right).
\label{sym-gauge-zeromode-wf}\end{equation}
Higher Landau level wave functions are derived as
$\ket{n, m} = \frac{(a^\dagger)^n}{\sqrt{n!}} \frac{(b^\dagger)^{n-m}}{\sqrt{(n-m)!}} \ket{0,0}$.
Thus, the coordinate representation of wave functions is given by
\begin{equation}
\phi_{n,m}(\tilde{\bm{r}}) = N_{m,n}
\exp\left(-\frac{|\tilde{\chi}|^2}{4} \right) \tilde{\chi}^{|m|} L_n^{(|m|)} \left( \frac{|\tilde{\chi}|^2}{2} \right),
\end{equation}
where
$N_{m,n} =  \frac{(-1)^n}{\sqrt{2 \pi} l_z} \sqrt{\frac{n!}{2^{(n - m)}(n - m)!}}$ is a normalization constant and
$L_n^{m}$ is Laguerre polynomial $L_n^m (t) = \frac{1}{n!} \e^t t^{-m} \frac{\diff^n}{\diff t^n}(\e^{-t} t^{n+m})$.

In particular, the $n = 0$ wave function is written by
\begin{equation}
  \label{symmetric-gauge-zeromode-wf2}
  \phi_{0,m}(\tilde{\bm{r}}) = \frac{1}{\sqrt{2 \pi 2^m m!} l_z}
  \tilde{\chi}^m \exp\left(-\frac{|\tilde{\chi}|^2}{4} \right).
\end{equation}
We show $|\phi_{0,0}(\tilde{\bm{r}})|^2$ and  $|\phi_{0,m}(\tilde{\bm{r}})|^2$ in \fref{fig:landau-level-wave-function}.
In the presence of Dirac-cone tilting, the energy contour of the cone becomes elliptic.
From the uncertainty principle, we expect that the wave function shrinks in the tilt direction.
In fact, for a tilted Dirac cone, the zero energy Landau level wave
function is anisotropic and shrinks in the tilt direction as shown in \fref{fig:landau-level-wave-function}.
For a non-tilted Dirac cone, the zero energy Landau level wave function is isotropic in real space (not shown).

\begin{figure}[htb]
\includegraphics[width=\linewidth]{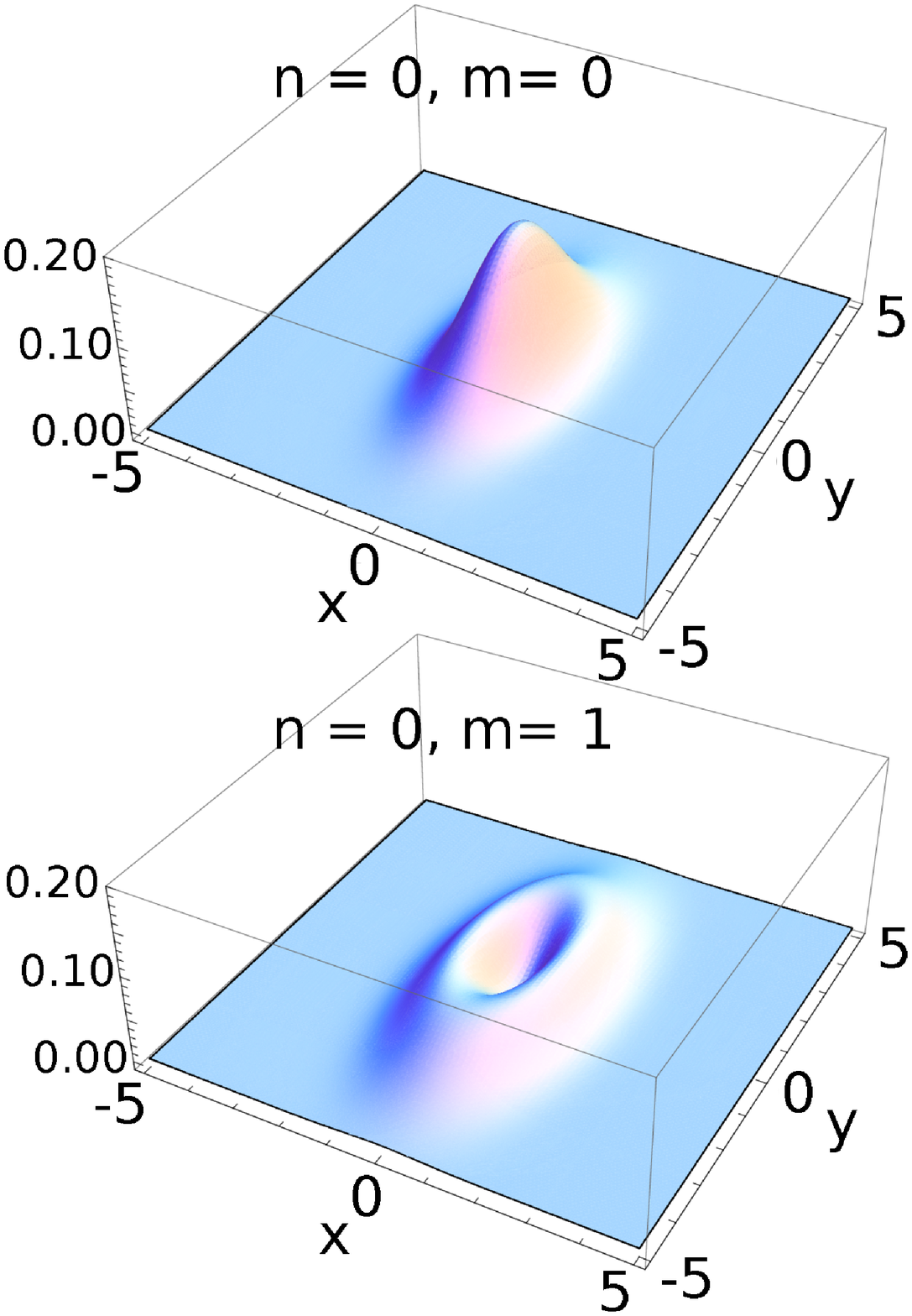}
\caption{Lowest Landau level wave functions $|\phi_{0,m}(\tilde{\bm{r}}) |^2$}
\label{fig:landau-level-wave-function}
\end{figure}

\subsection{Landau gauge case}
\label{landau-gauge-case}

In interlayer magnetoresistance calculation, it is convenient to take the Landau gauge.
We choose the gauge $A_x = B_y z + A_x^{(z)}$, $A_y = - B_x z + A_y^{(z)}$, $A_z = 0$, $\partial_x A_y^{(z)} - \partial_y A_x^{(z)} = B_z$
and
perform gauge transformation
\begin{equation}
  \psi = \exp \left[
    - \ii z \left( \frac{x}{l_y^2} - \frac{y}{l_x^2} \right)
  \right] \psi'.
  \label{landau-gauge-out-inplanem}
\end{equation}
Then we obtain
\begin{equation}
\pi_x = - \ii \hbar \partial_x - e A_x^{(z)}
,\quad
\pi_y = - \ii \hbar \partial_y - e A_y^{(z)}.
\end{equation}
We rescale and rotate the system as equation~\eref{rotate-coordinate}.
After this transformation, we take as $A_{x'}^{(z)} = - B_z y'$ and $A_{y'}^{(z)} = 0$.
The transformation of equation \eref{shift-dynamic-momentum} is equivalent to
$\tilde{x} = \frac{1}{\sqrt{\gamma}} x',
\
\tilde{y} = \sqrt{\gamma} \left(y' + \frac{\tau}{v} \frac{E \tilde{v}_0}{\gamma^2}\frac{1}{e B_z}\right)$.
We choose the operator $\tilde{Y}$ to also commute with Hamiltonian.
In this gauge, the operator $\tilde{Y}$ is given by $\tilde{Y} = l_z^2 \tilde{p}_x / \hbar$,
the momentum in the $\tilde{x}$ direction is conserved in this gauge.
Thus, the wave function is given by
\begin{equation}
\phi(\tilde{x},\tilde{y}) = \frac{1}{\sqrt{L_x}} \phi(\tilde{y}) \e^{\ii \tilde{k} \tilde{x}} = \frac{1}{\sqrt{L_x}} \phi(y) \e^{\ii k x},
\end{equation}
where $L_x$ is the length of the system in the $x$ direction.
The ladder-operator $\tilde{a}$ is given by
\begin{equation}
\tilde{a}
=
- \frac{1}{\sqrt{2} l_z}(\tilde{\eta} + l_z^2 \tilde{\partial}_\eta),
\end{equation}
where
\begin{equation}
\tilde{\eta} = \tilde{y} - \tilde{Y} = \sqrt{\gamma} \left(y' - l_z^2 k\right) - \sqrt{2 n} \tmag l_z.
\end{equation}
The wave function for zero-mode eigenfunction $\phi_0(y)$
is obtained by solving $\tilde{a} \phi_0(y) = 0$.
The eigenfunction is given by
\begin{equation}
\phi_n(\tilde{\eta})
= \frac{(-1)^n}{\pi^\frac{1}{4} \sqrt{2^n n! l_z}} \exp\left[ - \frac{\tilde{\eta}^2}{2 l_z^2} \right] H_n \left(\frac{\tilde{\eta}}{l_z}\right),
\end{equation}
where the Hermite polynomials is given by $H_n(x) = (-1)^n \e^{x^2} \frac{d^n}{d x^n} \e^{- x^2}$.
\section{Interlayer Magnetoresistance}
\label{interlayer-magneto-resistance}

Now we compute the interlayer magnetoresistance and discuss pressure effects on it.
We represent the magnetic field as $\bm{B} = (B_x, B_y, B_z) = B (\cos{\theta} \cos{\phi}, \cos{\theta} \sin{\phi}, \sin{\theta})$.

The interlayer tunneling between $j_z$-th plane and $(j_z+1)$-th plane is described by
\begin{equation}
  H_c = - t_c \sum_{j_z,\sigma=\uparrow,\downarrow} \int d^2r \, \hat{\psi}_{j_z,\sigma}(\bm{r})^\dagger \hat{\psi}_{j_z + 1,\sigma}(\bm{r}) + h.c.
\end{equation}
where $t_c$ is interlayer transfer energy.
In the Landau gauge, the momentum in the $\tilde{y}$ direction is conserved
because the central coordinate $\tilde{X}$ is conserved,
hence the operator $\hat{\psi}$ is written by
\begin{equation}
  \hat{\psi}_{j_z}(\bm{r}) = \sum_{n, X} \psi_{n, X} (x,y,j_z) \hat{c}_{X, j_z}.
\end{equation}
Using equation~\eref{landau-gauge-out-inplanem},
Landau wave function is given by
\begin{equation}\eqalign{
  \psi_{n, X}(x,y,z)
  = \frac{1}{\sqrt{L_x}} \exp \left[\ii z K \right] \psi'_{n, X}(y,z) \e^{\ii \tilde{k} \tilde{x}},
  \label{landau-gauge-wave-function}}
\end{equation}
where $K$ represents a phase factor defined by
\begin{equation}
  \eqalign{
  \fl K(x,y, \theta, \phi, \varphi) = \biggl[
  x (\sin \phi \cos \varphi - \alpha^{-2} \cos \phi \sin \varphi) \\
  - y (\alpha^2 \sin \phi \sin \varphi  + \cos \phi \cos \varphi) \biggl] \, \frac{e}{\hbar} B \cos{\theta} .}
\end{equation}
Here the current operator is written by
\begin{equation}\eqalign{
  \fl J_z
  =
  \ii e t_c \frac{1}{L_x} \int dx \int dy
  \sum_{n, n', k, k'} \Biggl[
  \e^{\ii (\tilde{k} - \tilde{k}')\tilde{y}}
  \e^{\ii a_c K(x,y, \theta, \phi, \varphi)}\\
  \times \psi'^\ast_{n',k'} \psi'_{n,k}
  \hat{c}^\dagger_{n, k, j_z}
  \hat{c}_{n', k', j_z + 1}
  + h.c. \Biggl],}
\end{equation}
where the $x$-integration gives $k' = k + \delta k$.
Hence, the center of mass $\tilde{Y}'$ is written as
$\tilde{Y}' = \tilde{Y} + l_z^2 \alpha \sqrt{\gamma} \delta k$,
where
\begin{equation}
  \delta k = \frac{e B}{\hbar} \cos{\theta} (\sin \phi \cos \varphi - \alpha^{-2} \cos \phi \sin \varphi) a_c.
\end{equation}
The current operator $J_z$ becomes
\begin{equation}\eqalign{
  J_z
  = 
  \frac{1}{L_x} \sum_{n, n', k, j_z} \Biggl[ &
  \int dy \exp\biggl\{- \ii a_c \frac{e}{\hbar} B \cos{\theta} (\alpha^2 \sin \phi \sin \varphi \\
    & \qquad + \cos \phi \cos \varphi) y\biggr\} \\
  & \times \phi^\dagger_{n',k + \delta k} \phi_{n,k}
  \hat{c}^\dagger_{n, k, j_z}
  \hat{c}_{n', k + \delta k, j_z + 1}
  + h.c. \Biggl].
\label{current_jz}}
\end{equation}

The matrix element $\braket{n,k,j_z|J_z|n',Y',j_z'}$ is written as
\begin{equation}\eqalign{
  \fl \braket{n,Y,j_z|J_z|n',Y',j_z'} \\
  = \ii e t (\delta_{j+1, j'} - \delta_{j-1,j'}) \delta_{\tilde{Y}', \tilde{Y} + l_z^2 \delta k}\\
  \quad \times \int\!dy
  \exp\left\{- \ii a_c \frac{e}{\hbar} B \cos{\theta} (\alpha^2 \sin \phi \sin \varphi  + \cos \phi \cos \varphi) y\right\} \\
  \quad \times
  \phi^\dagger_{n'} \left(\frac{\alpha \sqrt{\gamma}}{l_z} y - \tmag \sqrt{2 n'}\right)
  \phi_{n} \left(\frac{\alpha \sqrt{\gamma}}{l_z}(y - l_z^2 \delta k) - \tmag l_z \sqrt{2 n}\right).
  \label{matrix-element}}
\end{equation}
From the Kubo formula,
the interlayer magnetoresistance $\sigma_{zz}$ is given by
\begin{equation}
  \sigma_{zz}(\omega)
  =
  \frac{i}{\hbar} \sum_{\mathrm{spin}} \sum_{n,k,j \atop n',k',j'}
  \left[- \frac{f(E_{n'}) - f(E_n)}{E_{n'} - E_n}\right]
  \frac{|\braket{n,k,j|J_z|n'k'j'}|^2}{\hbar \omega + \ii \delta + (E_n - E_{n'})},
  \label{sigma_zz_kubo_formula}
\end{equation}
where the summations with respect to the layer index $j$ and the wave number $k$ yield
the number of layer $N_\mathrm{layer}$ and the Landau level degeneracy $\frac{1}{2 \pi l_z^2} = \frac{|e| B_z}{2 \pi \hbar}$, respectively.
The interlayer magnetoresistance $\rho_{zz}$ takes the form \cite{morinari_possible_2009} 
\begin{equation}
  \label{interlayer-magnetoresistance-zeromode}
  \frac{\rho_{zz}}{\rho_{zz}(B=0)}
  = 
  \frac{B_0}{
    B_0  + B \sin{\theta}
    \exp\left[
      -\frac{1}{2}
      \left( \frac{a_c}{l_z}  \right)^2
      \frac{\cos^2{\theta}}{\sin^2{\theta}} I(\phi,  \alpha, \tang, \tmparam)
    \right]},
\end{equation}
with
\begin{equation}\eqalign{
  I(\phi,  \alpha, \tang, \tmparam) =
  & \tmparam \left(\alpha \sin \phi \cos \tang - \frac{1}{\alpha} \cos \phi \sin \tang\right)^2 \\
  & +  \frac{1}{\tmparam} \left(\alpha \sin \phi \sin \tang  + \frac{1}{\alpha} \cos \phi \cos \tang\right)^2,
  \label{integral-of-magnetoresistance}}
\end{equation}
where $B_0$ is the resistance in the absence of a magnetic field.
This formula is derived by using the zero-mode Landau level wave function.
To justify this approximation, the magnetic field $B_z$ should be large enough or the temperature is low enough to satisfy the relation $E_1 > k_B T$.

The anisotropy of the Landau level wave function shown in \fref{fig:landau-level-wave-function}
leads to anisotropy
in the interlayer magnetoresistance.
\Fref{fig:physical_view} shows the physical picture of the dependence of interlayer magnetoresistance on the in-plane magnetic field direction.
The in-plane magnetic fields, $B_x$ and $B_y$, are treated by the gauge transformation \eref{landau-gauge-out-inplanem},
which gives rise to the phase factor when the electron hops between one layer to the adjacent layer.
\Fref{fig:physical_view} shows the case that the in-plane magnetic field is parallel to the $x$-axis.
In this case, the phase factor is written by $\exp\left(\ii a_c \frac{e B_x}{\hbar} y\right)$.
The wave function oscillates in real space
along the direction perpendicular to the in-plane magnetic field
because of the phase factor.
As a consequence, the matrix element \eref{matrix-element} is reduced
when the in-plane magnetic field is perpendicular
to the direction in which the wave function is extended.
Reflecting the real space anisotropy in the wave function,
the matrix element depends on the direction of the in-plane magnetic field, $\phi$.
The inter-layer magnetoresistance, thus, depends on $\phi$.

\begin{figure}[htb]
\centering
\includegraphics[width=0.5\linewidth]{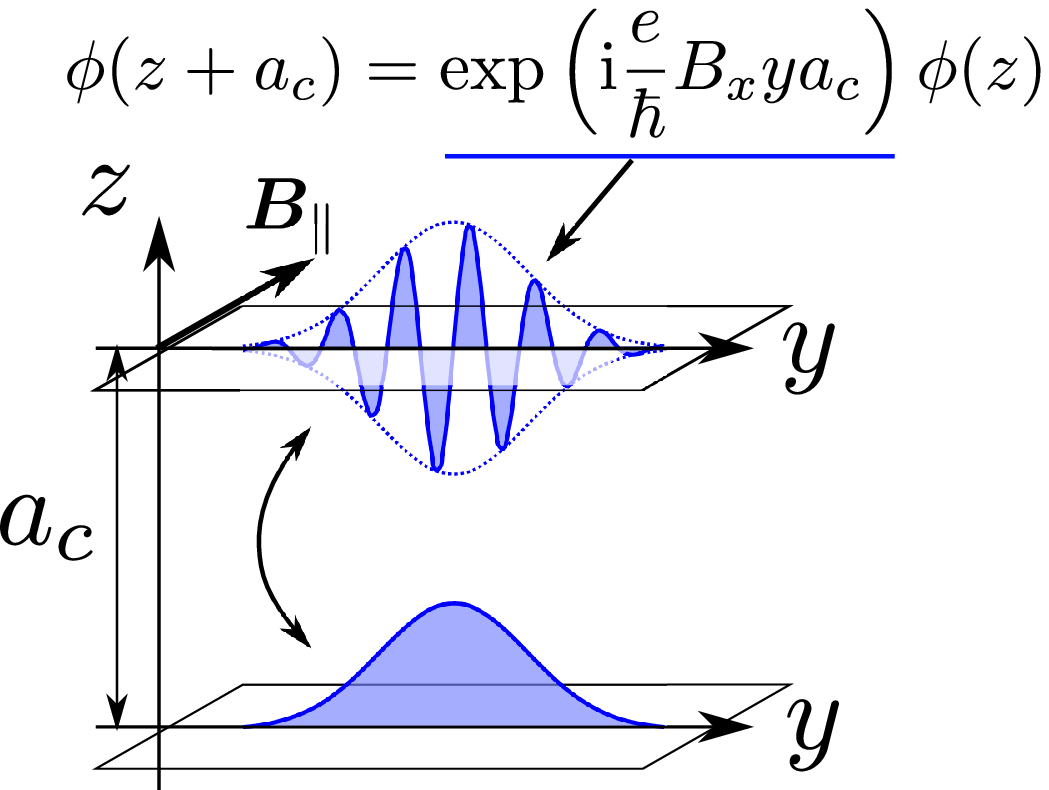}
\caption{Physical picture of the dependence of the matrix element
on the in-plane magnetic field direction.}
\label{fig:physical_view}
\end{figure}
\begin{figure}[htb]
\centering
\includegraphics[width=1.0\linewidth]{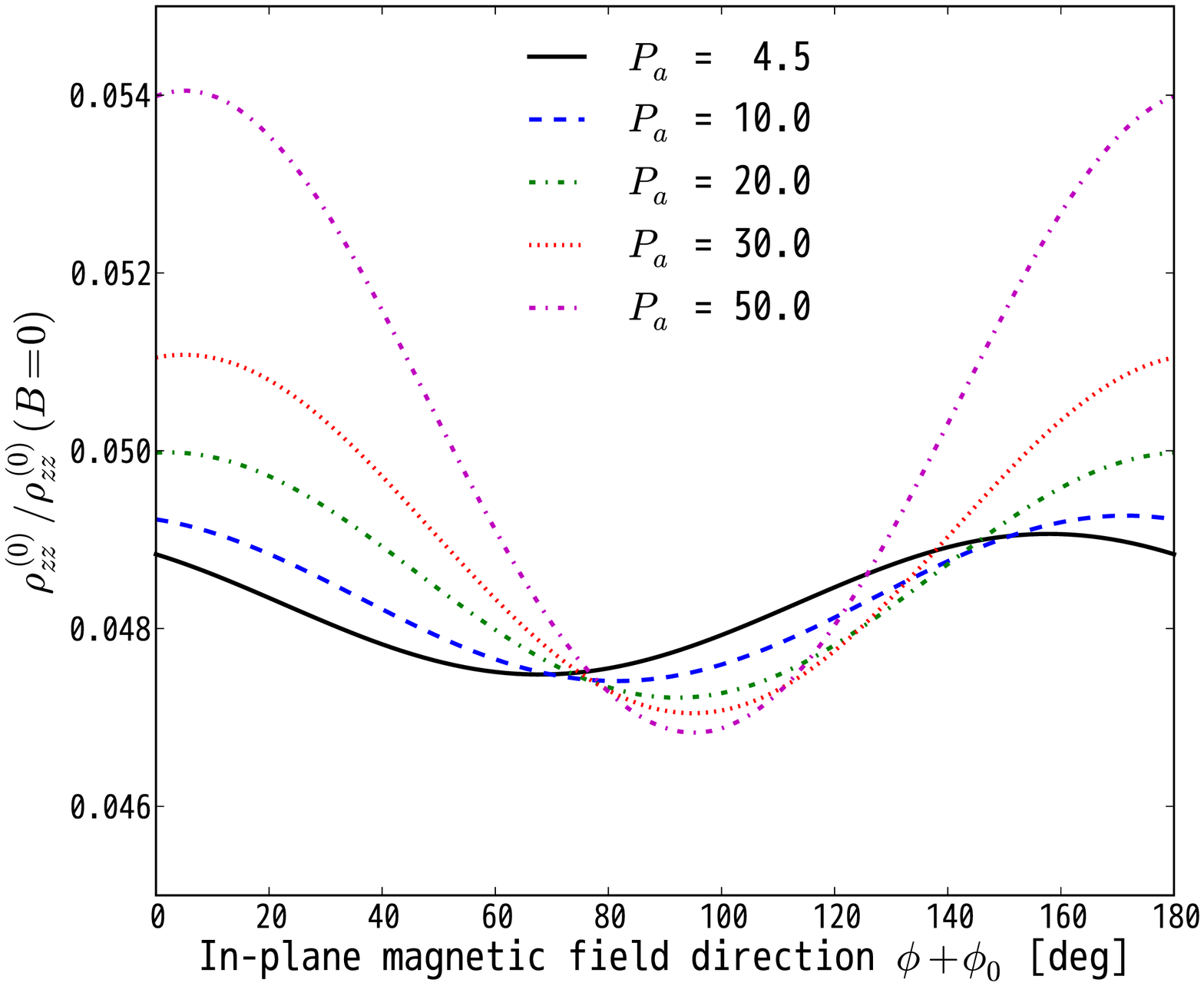}
\caption{The dependence of azimuthal angle dependence magnetic field direction on the interlayer magneto resistance for various values of uniaxial pressure.}
\label{fig:IMRvsPhi_pdep}
\end{figure}

\Fref{fig:IMRvsPhi_pdep} shows the in-plane magnetic field direction dependence of the interlayer magnetoresistance for different pressures.
When the pressure increases, the minimum of the magnetoresistance moves to 90 degrees and the peak of interlayer magnetoresistance increases.
In high pressure region, the parameter $\alpha$ increases as shown in \fref{fig:parameter}(a), so the effect from the anisotropy coming from non-tilting effect becomes dominant.
This growth results in the increase of the interlayer magnetoresistance peak.
In this case, the energy contour shrinks along the a$^\ast$-axis by the uniaxial pressure,
so the interlayer magnetoresistance takes the minimum when the in-plane magnetic field is parallel to a$^\ast$-axis, i.e., 90 degrees.

\section{Summary}
\label{summary}

In the present study, we examined pressure effects on Dirac fermions in \abedt within the tight-binding model.
The electron and valence bands are degenerate at two contact points $\bm{k}_0$ and $- \bm{k}_0$ in the Brillouin zone.
They are located at the Fermi level under wide pressure range.
The pressure dependence of the distance between contact points in the Brillouin zone also depends on the interaction parameters.
In the vicinity of the merging, ``valley spin'' picture would breakdown because the coupling between two valleys, which is usually neglected in graphene, becomes strong.
This merging behavior may be observed in the pressure range $40 \leq P_a < 70$ kbar, and the interaction parameter $x \leq 0.6$.
Around that pressure, we expect a rapid increase of the interlayer resistivity
coming from the opening of an energy gap.
This suggests that this system is useful for investigating valley splitting
effect that is still in controversial in graphene.
We show the exact solution of the Landau level on the tilted Weyl equation by using the symmetric and Landau gauges.
Because of the tilt, the Landau level wave functions become anisotropic and shrink in the tilt direction in real space.
We calculate the pressure dependence of the interlayer magnetoresistance by using the parameter estimated from the tight-binding model.
In high pressure region $P_a > 10$ kbar, anisotropy increases with pressure.
This increase results in the increase of the interlayer magnetoresistance peak.

\ack
This work was supported by the Grant-in-Aid for the Global COE Program "The Next Generation of Physics, Spun from Universality and Emergence" from the Ministry of Education, Culture, Sports, Science and Technology (MEXT) of Japan.
Numerical computations in ths work were carried out at the Yukawa Institute Computer Facility.
T.M. was supported by KAKENHI Grant No. 21740252.

\end{document}